\documentclass[aip,jcp,reprint,amsmath,amssymb,floatfix,citeautoscript]{revtex4-1}
\usepackage{cancel}
\usepackage{amsmath}
\usepackage{physics}
\usepackage{graphicx}
\usepackage{amssymb}
\usepackage{amsthm}
\usepackage{bm}
\usepackage{dcolumn}
\usepackage{braket}
\usepackage{longtable}

\usepackage{ragged2e}
\usepackage{txfonts}

\usepackage{color}
\usepackage[usenames,dvipsnames]{xcolor}
\definecolor{myblue}{rgb}{0,0,1}
\usepackage[breaklinks=true,colorlinks=true,linkcolor=myblue,urlcolor=myblue,citecolor=myblue]{hyperref}

\let\vr\undefined
\newcommand{\vr}{{\bm{r}}}

\newcommand{\vx}{{\bm{x}}}

\newcommand{\nocc}{N_{\mathrm{o}}}
\newcommand{\nvir}{N_{\mathrm{v}}}

\begin{document}

\title{
Improving MP2 band gaps with low-scaling approximations to EOM-CCSD 
}

\author{Malte F. Lange}
\affiliation{Department of Chemistry, Columbia University, New York, New York 10027 USA}

\author{Timothy C. Berkelbach}
\email{tim.berkelbach@gmail.com}
\affiliation{Department of Chemistry, Columbia University, New York, New York 10027 USA}
\affiliation{Center for Computational Quantum Physics, Flatiron Institute, New York, New York 10010 USA}

\begin{abstract}
Despite its reasonable accuracy for ground-state properties of semiconductors
and insulators, second-order M\o ller-Plesset perturbation theory (MP2)
significantly underestimates band gaps. Here, we evaluate the band gap predictions of partitioned
equation-of-motion MP2 (P-EOM-MP2), which is a second-order approximation to
equation-of-motion coupled-cluster theory with single and double excitations.
On a test set of elemental and binary semiconductors and insulators, we find that P-EOM-MP2
overestimates band gaps by 0.3~eV on average, which can be compared to the 
underestimation by 0.6~eV on average exhibited by the G$_0$W$_0$ approximation with a PBE reference.
We show that P-EOM-MP2, when interpreted as a Green's function-based theory, has a self-energy
that includes all first- and second-order diagrams and a few third-order diagrams.
We find that the GW approximation performs better for materials with small gaps and P-EOM-MP2
performs better for materials with large gaps, which we attribute to their superior treatment of
screening and exchange, respectively.
\end{abstract}

\maketitle

Second-order M\o ller-Plesset perturbation theory (MP2) is the simplest ab
initio treatment of dynamical electron correlation.  Its low cost makes it especially
attractive for large systems including periodic solids.  Although periodic MP2 has been
found to perform reasonably well for the description of ground-state 
properties~\cite{Ayala2001,Hirata2004,Hirata2009,Marsman2009,Gruneis2010,Maschio2010,Usvyat2010,Katouda2010,Goeltl2012,DelBen2012,DelBen2013,McClain2017},
its performance is less satisfactory for charged excitation energies and band
gaps~\cite{Gruneis2010,Iskakov2019}.  For example, in Ref.~\onlinecite{Gruneis2010}, MP2 was applied to thirteen 
semiconductors and insulators and exhibited average errors of 0.5\%
for lattice constants, 4.1\% for bulk moduli, and 0.23~eV for cohesive energies,
but predicted negative band gaps for
materials that are known to be semiconducting, such as silicon and silicon
carbide.  This unsatisfactory performance was attributed to the lack of
screening in finite-order perturbation theory.  Indeed, the GW
approximation~\cite{Hedin1965,Strinati1982,Hybertsen1986} and equation-of-motion 
coupled-cluster theory~\cite{Monkhorst1977,Nooijen1993,Stanton1994,Nooijen1995a,Lange2018}
describe excitation energies with infinite-order perturbation theory and predict
accurate band gaps of semiconductors~\cite{Hybertsen1986,Schilfgaarde2006,McClain2017}, albeit with a computational cost
that is higher than that of MP2.

Here, we study the performance of a second-order approximation to equation-of-motion
coupled-cluster theory with single and double excitations (EOM-CCSD), first presented
in Refs.~\onlinecite{Nooijen1995,Gwaltney1996}.  Despite making
sequential second-order approximations, the method will be seen to be equivalent
to the use of a self-energy containing all second-order diagrams and a few third-order diagrams.  

Consider the M\o ller-Plesset partitioning of the many-body Hamiltonian, leading to the
Hartree-Fock (HF) orbitals $\phi_p(\vr)$ with orbital energies $\varepsilon_p$; as usual,
we denote the orbitals occupied in the HF determinant by $i,j,k,l$, those unoccupied by
$a,b,c,d$, and general orbitals by $p,q,r,s$.
The self-energy evaluated to second-order in perturbation theory is
\begin{equation}
\label{eq:mp2}
\Sigma_{pq}^{\mathrm{MP2}}(\omega) = \frac{1}{2}\sum_{lcd}\frac{V_{pl}^{cd}V_{cd}^{ql}}{\omega+\varepsilon_l-\varepsilon_c-\varepsilon_d}
    + \frac{1}{2}\sum_{kld}\frac{V_{pd}^{kl}V_{kl}^{qd}}{\omega+\varepsilon_d-\varepsilon_k-\varepsilon_l},
\end{equation}
where the antisymmetrized two-electron integrals are defined by
$V_{pq}^{rs} = \langle pq|rs\rangle - \langle pq|sr\rangle$, with
\begin{equation}
\langle pq|rs \rangle 
    = \int d\vx_1 d\vx_2\phi^*_{p}(\vx_1)\phi^*_{q}(\vx_2)r_{12}^{-1}\phi_{r}(\vx_1)\phi_{s}(\vx_2)
\end{equation}
and $\vx$ is a combined space and spin variable.
Unlike the GW approximation, the MP2 self-energy has exact second-order exchange and is therefore
free of self-screening error.

\begin{figure}[b]
\centering
\includegraphics[scale=1.0]{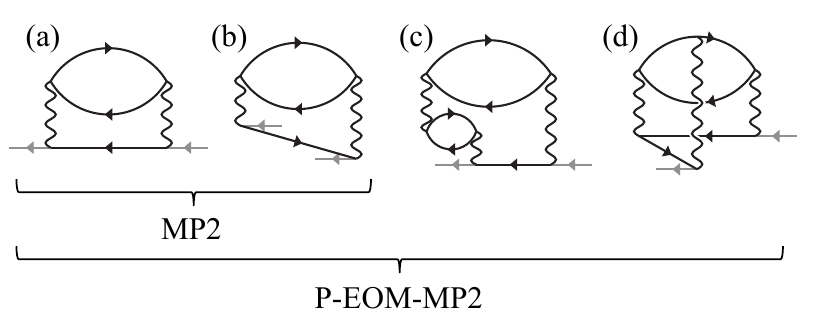}
\caption{
Self-energy diagrams included in the MP2 and P-EOM-MP2 Green's function beyond first-order.  All diagrams
are time-ordered with time increasing from left to right; hole lines point towards decreasing
time and particle lines point towards increasing time.  All Coulomb interactions (wavy lines) are
antisymmetrized, yielding exchange diagrams not explicitly drawn here.  The MP2 self-energy
includes diagrams (a) and (b) only.  The P-EOM-MP2 self-energy includes all four diagrams shown.
The GW self-energy includes the non-exchange versions of diagrams (a), (b), (c), and many others.
}
\label{fig:diagrams}
\end{figure}

An alternative theory can be obtained by a second-order approximation to EOM-CCSD,
leading to a method originally referred to as EOM-MBPT(2)~\cite{Nooijen1995} or EOM-CCSD(2)~\cite{Stanton1995}.
In this method, the ground-state CCSD amplitudes
are approximated by their MP2 values, avoiding the expensive iterative solution of the CCSD amplitude
equations.  In this work, we consider the additional approximation of partitioning the EOM Hamiltonian
into single and double excitation spaces and perturbatively treating the latter. Under this approximation,
the large double excitation block of the similarity-transformed Hamiltonian is a diagonal matrix
of orbital energy differences. This method has been
referred to as DSO-GF~\cite{Nooijen1995} and P-EOM-MBPT(2)~\cite{Gwaltney1996,Goings2014}; because
we always use a Hartree-Fock reference, we will refer to the method as P-EOM-MP2.

Unlike typical Green's function techniques, the EOM approach yields ionization potentials (IPs) and
electron affinities (EAs) from separate eigenvalue calculations.
In practice, these eigenvalues are found iteratively using the Davidson algorithm.
As shown by Nooijen
and Snijders~\cite{Nooijen1995}, the P-EOM-MP2 IPs can equivalently be
obtained from the self-consistent eigenvalues of a matrix with elements
$\varepsilon_{i}\delta_{ij} + \Sigma_{ij}^{\mathrm{EOM}}(\omega)$, where
\begin{equation}
\label{eq:eom}
\Sigma_{ij}^{\mathrm{EOM}}(\omega) 
    = \frac{1}{2}\sum_{lcd}\frac{V_{il}^{cd}V_{cd}^{jl}}{\varepsilon_j+\varepsilon_l-\varepsilon_c-\varepsilon_d} 
    + \frac{1}{2}\sum_{kld}\frac{W_{id}^{kl}V_{kl}^{jd}}{\omega+\varepsilon_d-\varepsilon_k-\varepsilon_l}
\end{equation}
and likewise for the EAs.
The above matrix is clearly similar to the MP2 self-energy matrix~(\ref{eq:mp2}), except for three differences.
The first difference is the neglected coupling between the particle and hole
spaces. Within the common diagonal approximation to the self-energy, this
coupling is irrelevant and we have numerically confirmed that it is a negligible
difference in this work.
The second difference is the perturbative 
replacement of $\omega=\varepsilon_j$ in one of the two terms.  When this replacement is done in the
MP2 self-energy, we find that it makes the results slightly \textit{worse} and is thus not responsible
for the improvement to be shown in the P-EOM-MP2 band gaps (vide infra).

The third and most important difference is the presence of 
the intermediate
\begin{equation}
\label{eq:W}
W_{id}^{kl} = V_{id}^{kl} + P_-(kl)\sum_{me} V_{im}^{ke} t_{lm}^{de}
    + \frac{1}{2} \sum_{ef} V_{id}^{ef} t_{kl}^{ef}
\end{equation}
where the antisymmetrization operator is $P_-(kl) A_{kl} = A_{kl} - A_{lk}$
and
\begin{equation}
t_{ij}^{ab} = \frac{V_{ab}^{ij}}
    {\varepsilon_i+\varepsilon_j-\varepsilon_a-\varepsilon_b}.
\end{equation}
Viewed in terms of the similarity-transformed Hamiltonian $\bar{H} =
e^{-T}He^{T}$, the first term in Eq.~(\ref{eq:eom}) reflects a renormalization
of the one-body interactions due to ground-state correlation and the presence of
$W$ in the second term reflects a renormalization of the two-body interactions,
i.e.~it is a screened Coulomb interaction.
Alternatively, the intermediate $W$ can be understood as the inclusion of a few third-order
self-energy diagrams, with a perturbative evaluation of the frequency denominator.
In Fig.~\ref{fig:diagrams}, we show the time-ordered self-energy diagrams included in the
MP2 and P-EOM-MP2 Green's functions beyond first-order (i.e.~beyond Hartree-Fock).

Assuming $\nocc$ occupied orbitals and $\nvir$ virtual orbitals with $\nvir >
\nocc$, then the more expensive electron affinity EOM-CCSD (EA-EOM-CCSD) has an
iterative $O(\nocc^2\nvir^4)$ cost due to ground-state CCSD and an iterative
$O(\nocc^2\nvir^3)$ cost per eigenvalue for excited-state matrix-vector multiplication.  The
P-EOM-MP2 method reduces the above to a non-iterative $O(\nocc^2\nvir^2)$ cost
due to ground-state MP2 (ignoring the integral transformation) and an iterative
$O(\nocc\nvir^3)$ cost per eigenvalue due to excited-state matrix-vector multiplication.
This significant cost reduction makes P-EOM-MP2 an attractive approach for complex
materials.
(Strictly speaking, P-EOM-MP2 has a non-iterative $O(N^6)$ step due to the formation
of the intermediate~(\ref{eq:W}), but this is typically not the most time-consuming step.
If necessary, the one-time intermediate construction can be avoided but results in
an iterative $O(N^5)$ cost.)

We have applied the above two theories to the calculation of the minimum band
gaps of eleven simple, three-dimensional semiconductors and insulators.
Seven have a diamond/zinc-blende crystal structure: Si, 
SiC, GaP, BP, GaN, C, BN; two have a rock-salt crystal structure: MgO and LiF; 
and two have a face-center cubic crystal structure: Ar and Ne. 
Calculations were done
with periodic boundary conditions using the PySCF software
package~\cite{Sun2017,Sun2020}. All calculations were done without
pseudopotentials using the all-electron cc-pVTZ basis set except for Ne and Ar,
which used the aug-cc-pVTZ basis set.  
Calculations using larger basis sets (not shown) suggest that our results are close to the
basis set limit, consistent with analogous results obtained with the GW
approximation~\cite{Zhu2021}.  Two-electron integrals were calculated by
periodic Gaussian density fitting~\cite{Sun2017a} using JKFIT
auxiliary basis sets~\cite{Weigend2002}.

For charged excitation energies, finite-size effects are large~\cite{McClain2017,Yang2020}.
Here, we have included one Madelung constant correction to the occupied
orbital energies and another to the final IPs. The former correction has no impact in 
wavefunction-based theories such as EOM-CCSD, but does have an impact in finite-order
perturbation theories (similar to the differing behaviors of ground-state CCSD and MP2);
the latter correction is familiar from periodic calculations of charged systems and can
be given a many-body interpretation on the basis of the excited-state structure
factor~\cite{Yang2020}.
We have performed calculations with $N_k=2^3 - 5^3$ $k$-points
sampled uniformly in the Brillouin zone.  Band gaps were then extrapolated to
the thermodynamic limit assuming an $O(N_k^{-1/3})$ finite-size error.  Other
treatments of finite-size effects are possible, but all are expected to exhibit
finite-size errors with the same scaling.

\begin{figure}[t]
\centering
\includegraphics[scale=0.9]{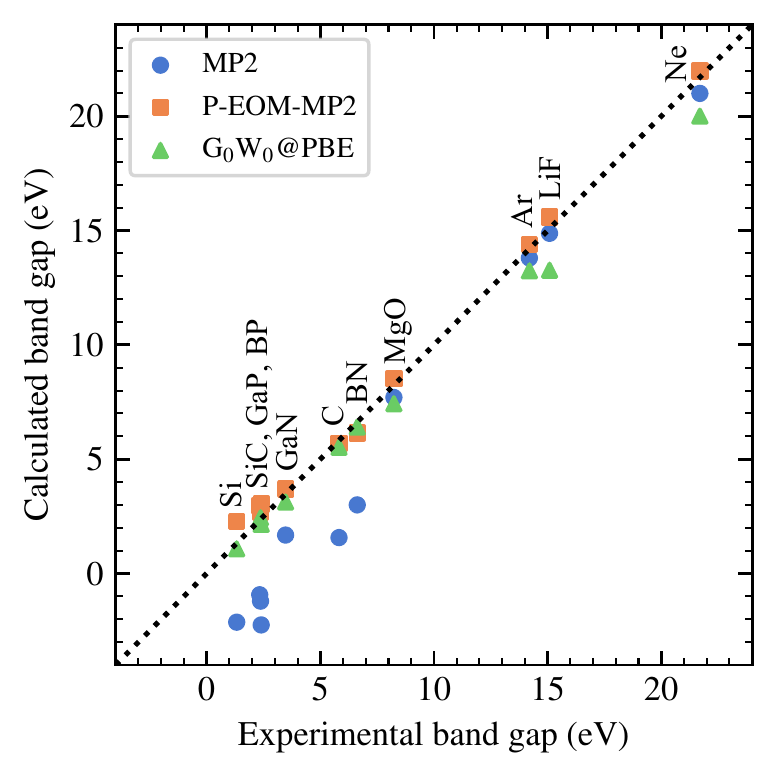}
\caption{
Comparison of calculated band gaps to experimental band gaps (including
zero-point renormalization) for the eleven semiconducting and insulating
materials indicated.  GW approximation results were obtained at the
G$_0$W$_0$@PBE level of theory.
}
\label{fig:exp_vs_pred}
\end{figure}

\begin{table*}[t]
  \centering
  \setlength{\tabcolsep}{10pt}
  \begin{tabular*}{0.98\textwidth}{@{\extracolsep{\fill}} lddddddddd}
\hline\hline
		Material & \multicolumn{1}{c}{$a$ (\r{A})} & \multicolumn{2}{c}{Reference} & \multicolumn{2}{c}{MP2} & \multicolumn{2}{c}{P-EOM-MP2} & \multicolumn{2}{c}{G$_0$W$_0$@PBE} \\
		& & \multicolumn{1}{c}{Expt. $E_\mathrm{g}$} & \multicolumn{1}{c}{el-ph} & \multicolumn{1}{c}{$E_\mathrm{g}$} & \multicolumn{1}{c}{Error} & \multicolumn{1}{c}{$E_\mathrm{g}$} & \multicolumn{1}{c}{Error} & \multicolumn{1}{c}{$E_\mathrm{g}$} & \multicolumn{1}{c}{Error} \\
		\hline
		Si    &  5.431 & 1.24 & -0.06 & -2.13 & -3.43 & 2.26  & 0.96  & 1.08  & -0.22 \\
		SiC   &  4.350 & 2.2  & -0.17 & -1.21 & -3.58 & 2.66  & 0.29 & 2.44  & 0.07 \\
		GaP   &  5.450 & 2.27 & -0.07 & -0.93 & -3.27 & 2.96  & 0.62  & 2.33  & -0.01  \\
		BP    &  4.538 & 2.4  &  -    & -2.25 & (-4.65) & 3.04 & (0.64) & 2.15 & (-0.25) \\
		GaN   &  4.520 & 3.30 & -0.18 & 1.68  & -1.80 &  3.70 &  0.22 & 3.13  & -0.35  \\
		C     &  3.567 & 5.48 & -0.33 & 1.57  & -4.24 & 5.70  & -0.11 & 5.52  & -0.29 \\
		BN    &  3.615 & 6.2  & -0.41 & 3.00  & -3.61 & 6.15  & -0.46 & 6.41  & -0.20 \\
		MgO   &  4.213 & 7.67 & -0.52 & 7.70  & -0.49 & 8.52  & 0.33  & 7.43  & -0.76 \\ 
	        Ar    &  5.256 & 14.2 & \sim 0& 13.80 & -0.40 & 14.38 &  0.18 & 13.24 & -0.96 \\
		LiF   &  4.035 & 14.5 & -0.59 & 14.88 & -0.21 & 15.59 & 0.50  & 13.27 & -1.82 \\
		Ne    &  4.429 & 21.7 & \sim 0& 21.00 & -0.70 & 21.98 &  0.28 & 20.01 & -1.69 \\
\hline
                MSE (eV)  &    &      &         &       & -2.17 &       & +0.28 &       & -0.62  \\
                MUE (eV)  &    &      &         &       &  2.17 &       &  0.40 &       &  0.64  \\
\hline\hline
  \end{tabular*}
  \caption{Minimum band gap $E_\mathrm{g}$ as measured experimentally and as
predicted by MP2, P-EOM-MP2 and G$_0$W$_0$@PBE (from Ref.~\onlinecite{Zhu2021}).
Errors in predicted band gaps are calculated with respect to experimental values
with electron-phonon (el-ph) renormalization.  All energies are in eV.  Mean signed 
error (MSE) and mean unsigned error (MUE) are given in eV.
percentage. Results on BP were excluded from error statistics due to the missing electron-phonon
renormalization.  Experimental band gaps are from
Refs.~\onlinecite{Madelung2004,Chiang1989,Schwentner1975}, zero-point contributions to electron-phonon
renormalization are from
Refs.~\onlinecite{Miglio2020,Antonius2015}, the thermal contribution to electron-phonon
renormalization for LiF is from 
Ref.~\onlinecite{Monserrat2016}, and G$_0$W$_0$@PBE results are from
Refs.~\onlinecite{Zhu2021,Shishkin2007}.  
}
  \label{tab:1}
\end{table*}

In Fig.~\ref{fig:exp_vs_pred}, we compare the minimum band gaps obtained by MP2, P-EOM-MP2, and
the GW approximation to experimental values at 300~K. Because the calculations do not
account for vibrational effects, we have adjusted the experimental values
according to calculated electron-phonon renormalizations from the literature~\cite{Miglio2020,Antonius2015}
based on the Allen-Heine-Cardona framework~\cite{Allen1976,Allen1981,Giustino2010}.
We only include the zero-point renormalization for all materials except LiF, which has a sizable thermal contribution
to the renormalization at 300~K~\cite{Monserrat2016}; for the other materials, this latter contribution is relatively small.
Lattice expansion is already accounted for because our lattice constants are experimental 300~K values.
Precise numbers and crystal geometries are given in Tab.~\ref{tab:1}.
We note that experimental band gaps and calculated electron-phonon renormalizations vary throughout the
literature.

Consistent with Ref.~\onlinecite{Gruneis2010}, we find that MP2 systematically
underestimates the band gap and predicts negative band gaps for Si, GaP, BP, and
SiC (our MP2 band gaps are similar to those of Ref.~\onlinecite{Gruneis2010}, but some differ
by as much as 0.5~eV, which we attribute to differences in the treatment of core electrons, basis set effects, 
finite-size effects).
Remarkably, the P-EOM-MP2 band gaps are a significant improvement and show
good agreement for all materials.  The mean signed error (MSE) is $+0.28$~eV and the mean
unsigned error (MUE) is 0.40~eV. The largest signed error is for Si ($+0.96$~eV),
which has the smallest gap of all materials considered. 

In Fig.~\ref{fig:exp_vs_pred} and Tab.~\ref{tab:1}, we also compare results to those 
calculated by the G$_0$W$_0$ approximation with a
PBE reference.  For all materials except LiF, we compare to all-electron, full-frequency 
calculations by Zhu and Chan~\cite{Zhu2021}, which were performed with PySCF using identical
treatments of core electrons and identical Gaussian basis sets. The result for LiF
is from Ref.~\onlinecite{Shishkin2007}.
Remarkably, the P-EOM-MP2 and GW approximation perform similarly well, despite
their underlying physical differences.  Roughly speaking, the GW approximation performs better for materials
with the smallest gaps while P-EOM-MP2 performs better for those with the largest gaps. 
The largest errors for the GW approximation are for the large-gap insulators, whose 
band gaps are underestimated by about 1~eV or more, which we attribute to the use
of a PBE starting point and the absence of second-order exchange. 
For the GW approximation,
the MSE is $-0.62$~eV and the MUE is $0.64$~eV. 

\begin{figure}[b]
\centering
\includegraphics[scale=0.95]{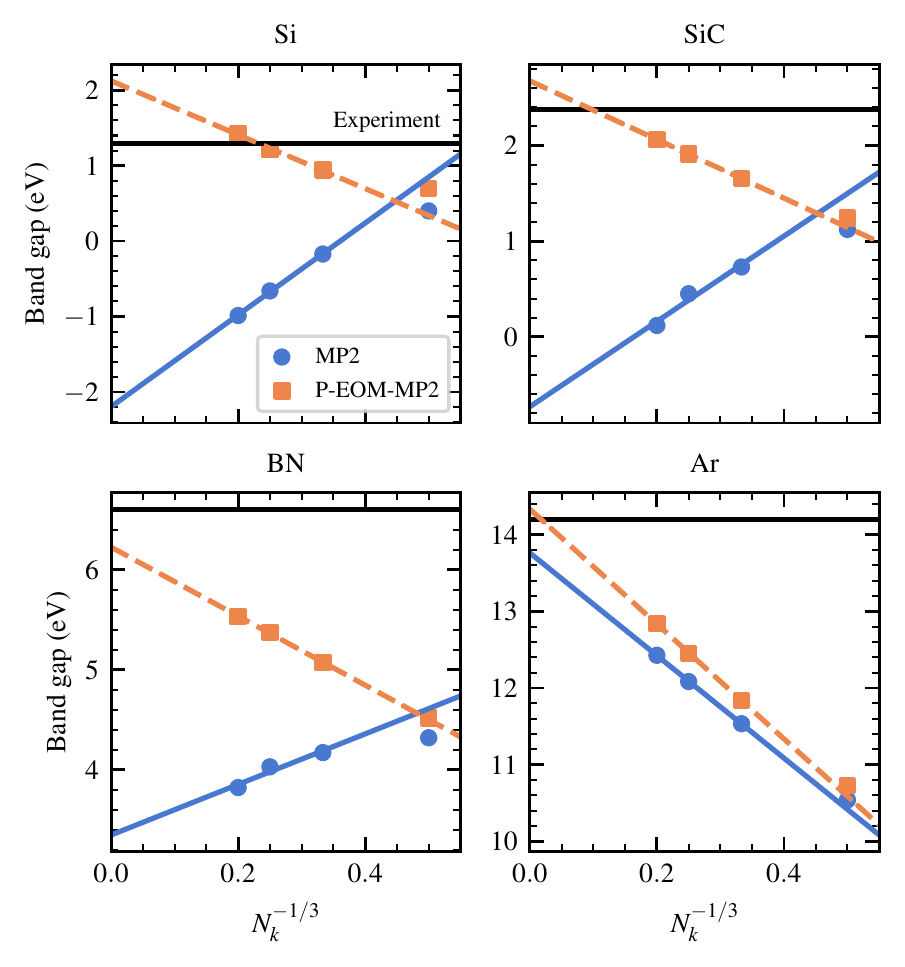}
\caption{
Convergence and extrapolation to the thermodynamic limit of the MP2 and
P-EOM-MP2 band gaps for four example materials. Experimental values are
corrected with a zero-point renormalization.
}
\label{fig:bandgaps}
\end{figure}

We performed additional calculations to estimate the effect of the various diagrams included in
the P-EOM-MP2 self-energy shown in Fig.~\ref{fig:diagrams}. The HF band gap is always much too large
and the direct second-order diagrams yield a large negative correction, which is largely responsible
for the performance exhibited by MP2. Consistent with Ref.~\onlinecite{Gruneis2010},
we find that the second-order exchange diagrams make a small contribution ($\lesssim 0.2$~eV) for small-gap
materials but a larger contribution ($\gtrsim 0.5$~eV) for large-gap materials. This
can be attributed to the more localized nature of the electronic states in large-gap insulators.
The screening diagram in Fig.~\ref{fig:diagrams}(c) makes a large contribution (1~eV or more) for all materials
and is most responsible for the significant improvement of P-EOM-MP2 over MP2. The final diagram
in Fig.~\ref{fig:diagrams}(d), a vertex correction beyond the GW approximation, typically raises the gap by
about $0.2$~eV. 

In Fig.~\ref{fig:bandgaps}, we show the convergence of the band gap towards the thermodynamic limit for
four of the materials considered here.  As mentioned earlier, the finite-size error is large and must be
removed by extrapolation. 
Interestingly, although MP2 and P-EOM-MP2 
give similar band gaps for small $k$-point meshes, they exhibit very different convergence to the thermodynamic
limit.  This difference is largest for materials with small band gaps.  We attribute this behavior
to our use of Madelung constant corrections in the HF orbital energies.  These corrections cause the HF gap
to converge to the thermodynamic limit from above, such that the systems with smaller $k$-point meshes are
more weakly correlated and the importance of third-order diagrams in the self-energy is diminished. On approach
to the thermodynamic limit, the system becomes more strongly correlated and the results of the two methods deviate.

In conclusion, we have shown that the P-EOM-MP2 approach, a second-order approximation to EOM-CCSD, 
yields solid-state band gaps that are a significant improvement
over those predicted by MP2.  The success of P-EOM-MP2 contradicts the
conventional wisdom that infinite-order screening is necessary for quantitative accuracy in band gap prediction.
Rather, P-EOM-MP2 represents an affordable balance of low-order screening and exchange, yielding
semiquantitative accuracy for materials with a wide range of band gaps. 
By starting from Hartree-Fock theory and including antisymmetrization in all interaction vertices, the method
is completely ab initio and free of self-interaction and self-screening errors.
We note that P-EOM-MP2 is very similar to CC2~\cite{Christiansen1995} and we therefore expect similar performance from
the latter, which also includes some amount of orbital relaxation.
Although P-EOM-MP2 has been found to perform well for three-dimensional materials, it will be interesting to
apply it to low-dimensional semiconductors, where screening is more complicated. 

\begin{acknowledgments}
This work was supported in part by the National Science Foundation under Grant
No.~CHE-1848369.  M.F.L.~was supported in part by the National Science
Foundation Graduate Research Fellowship Grant DGE-1746045.  We acknowledge
computing resources from Columbia University's Shared Research Computing
Facility project, which is supported by NIH Research Facility Improvement Grant
1G20RR030893-01, and associated funds from the New York State Empire State
Development, Division of Science Technology and Innovation (NYSTAR) Contract
C090171, both awarded April 15, 2010.  We also acknowledge computing resources
from the Flatiron Institute.  The Flatiron Institute is a division of the Simons
Foundation.
\end{acknowledgments}

\end{document}